\documentstyle[12pt]{article}
\begin{document}

\begin{titlepage}
\hfill{UM-P-97/07}

\hfill{RCHEP-97/01}

\vskip 4.2 cm

\centerline{{\large \bf 
Effects for atmospheric neutrino experiments}}
\centerline{{\large \bf
from electron neutrino oscillations}}
\vskip 0.2cm

\vskip 1.3 cm
\centerline{J. Bunn, R. Foot and R. R. Volkas}
\vskip 1.0 cm
\noindent
\centerline{{\it School of Physics,}}
\centerline{{\it Research Centre for High Energy Physics,}}
\centerline{{\it The University of Melbourne,}}
\centerline{{\it Parkville, 3052 Australia. }}

\vskip 1.0cm

\centerline{Abstract}
\vskip 0.7cm
\noindent
The minimal interpretation of the atmospheric
neutrino data suggests that the muon
neutrino oscillates into another species with a 
mixing angle close to the maximal $\pi/4$.
In the Exact Parity Symmetric Model, both 
the muon and electron neutrinos are expected to be 
maximally mixed with essentially sterile partners
($\nu'_{\mu}$ and $\nu'_e$ respectively).
We examine the impact of maximal $\nu_e - \nu'_e$ oscillations
on the atmospheric neutrino experiments. We estimate that
maximal $\nu_e - \nu'_e$ oscillations will have effects on 
atmospheric neutrino data 
for $|\delta m^2 \ (\nu_e - \nu_e')|
> 7 \times 10^{-5} \ eV^2$. 
For $\delta m^2$ in this range,
a slight but distinctive rise in the ratio of muon-like 
to electron-like
events is predicted for the low-energy sample. Furthermore, 
the ratio of low-energy electron-like events
with zenith angles less than $90\deg$ to those with
zenith angles greater than $90\deg$ should be greater than 1.

\end{titlepage}
There are three main experimental indications that neutrinos
have mass and oscillate: the atmospheric
neutrino anomaly\cite{kam,imb,sou}, the solar neutrino 
problem\cite{sn} and the LSND experiment\cite{lsnd}.
In the atmospheric neutrino experiments, the fluxes of 
electron and muon neutrinos resulting from cosmic ray 
interactions with the atmosphere are measured.
The results of these experiments are usually expressed
in terms of the quantity $R$, where
\begin{equation}
R \equiv {(n_{\mu}/n_e)_{data} \over (n_{\mu}/n_e)_{MC}}
\label{R}
\end{equation}
and $n_{\mu}$ and $n_e$ are the numbers of $\nu_{\mu}$ and 
$\nu_e$ induced events respectively. The ratio
$(n_{\mu}/n_{e})_{MC}$ is obtained from Monte Carlo simulated 
events based on
theoretical calculations of the fluxes\cite{nf}.
The experimental results are summarized in Table 1 for the low
energy (sub GeV) data. Observe that while no anomaly has been
observed by the Frejus\cite{frejus} or Nusex 
collaborations\cite{nusex},
the errors for these experiments are significantly larger
than the errors quoted by the Kamiokande\cite{kam} and IMB\cite{imb}
collaborations.  In this paper we assume that the anomaly is real 
and represents evidence for neutrino oscillations.

In this context, the atmospheric neutrino anomaly suggests that 
the muon neutrino oscillates maximally (or close to maximally) 
with another as yet unidentified species\cite{kam}.  
The solar neutrino problem\cite{sn}, on 
the other hand suggests that either the electron 
neutrino oscillates maximally (or near maximally) 
with another species 
\cite{gp}\cite{fv}\cite{acker}\cite{har}\cite{conf}\cite{js} 
{\it or} there are small angle 
oscillations which are enhanced due to matter effects 
in the sun\cite{msw}.
It is tempting to assume that both the 
atmospheric and solar neutrino 
anomalies are due to essentially the 
same mechanism, which then suggests
that they are both solved by large angle or 
maximal neutrino oscillations.

 From a model building perspective, several 
simple ideas have been put forward
to explain both the atmospheric and solar neutrino problems
via large angle oscillations, including,
\vskip 0.5cm
\noindent
(i) The electron and muon neutrinos oscillate into each other
with near maximal mixing\cite{acker}. This solution is also
compatible with the LSND experiment\cite{ackpak}. 
\vskip 0.5cm
\noindent
(ii) The three known neutrinos are maximally mixed 
with each other\cite{har, giu}.
This solution can explain the atmospheric and solar
neutrino problems but is incompatible with LSND.
\vskip 0.5cm
\noindent
(iii) All three of the neutrinos are maximally mixed with
a sterile species (we denote the three sterile neutrinos
by the notation $\nu'_e$, $\nu'_{\mu}$, 
$\nu'_{\tau}$)\cite{fv,flv,oth}.
This scenario is also compatible with the LSND experiment.
\vskip 0.5cm
\noindent
The purpose of this paper is to study some implications of
this last possibility for current atmospheric neutrino
experiments.

Although the solution (iii) is non-minimal, it can be theoretically
well motivated. Our interest in the above scheme comes from
the observation that it is naturally realized in gauge
models with exact (i.e. unbroken) parity symmetry\cite{flv}.
Exact parity symmetry is possible if the particle
content of the standard model is doubled (the doubling occurs
because each of the known particles has a mirror image which is a 
distinct particle).  If neutrinos have mass, and mass mixing
between ordinary and mirror neutrinos occurs, then
each of the known neutrinos is  necessarily
a maximal mixture of two states (assuming that
the mixing between generations is small)\cite{flv}.
There are also other interesting schemes which suggest
that the ordinary neutrinos oscillate maximally into
sterile states\cite{oth}. 

Maximal $\nu_e - \nu'_e$ oscillations 
reduce the flux of solar neutrinos by an energy
independent factor of 2 for the large range of parameters
\begin{equation}
3 \times 10^{-10} \stackrel{<}{\sim}
|\delta m^2_{ee'}|/eV^2 \stackrel{<}{\sim} 7.5 \times 10^{-3},
\label{1}
\end{equation}
where the upper bound is the most recent 
experimental bound\cite{exp}.
(Note that the MSW matter effects due to neutrino propagation
through the sun can be ignored if the electron
neutrino oscillates maximally\cite{fv}).
Maximal $\nu_{\mu} - \nu'_{\mu}$ oscillations
can explain the atmospheric neutrino anomaly 
provided that $10^{-3} \stackrel{<}{\sim} 
|\delta m^2_{\mu \mu'}|/eV^2 
\stackrel{<}{\sim} 10^{-1}$\cite{kam, barpak, cos}.
The best fit (obtained from a fit to the zenith
angle dependent multi-GeV neutrino data)\cite{kam} 
occurs for 
$|\delta m^2_{\mu \mu'}| \simeq 1.6\times 10^{-2} \ eV^2$.
Note however that the atmospheric neutrino experiments are 
sensitive to both $\nu_{\mu} -\nu'_{\mu}$ and $\nu_e - \nu'_e$
oscillations in principle. Assuming that the oscillations
are exactly maximal, we will study the constraints on the
parameters $\delta m^2_{ee'}$ and $\delta m^2_{\mu \mu'}$
suggested by the existing atmospheric neutrino experiments.

For maximal $\nu_{\alpha} - \nu_{\alpha}'$ oscillations 
(with $\alpha = e$ for $\nu_e - \nu'_e$ oscillations and
$\alpha = \mu$ for $\nu_{\mu} - \nu'_{\mu}$ oscillations), the 
probability that a weak eigenstate
neutrino $\nu_{\alpha}$ of energy $E_{\nu}$ remains  a 
weak eigenstate $\nu_{\alpha}$ after travelling a 
distance $L$ is in general\cite{msw},
\begin{equation}
P(\nu_{\alpha} \to \nu_{\alpha}, L, E_{\nu}) = 
1 - {\sin^2 \delta_m \over f},
\label{P}
\end{equation}
where $\delta_m$ is given by
\begin{equation}
\delta_m \simeq 1.27\left[{\delta m^2_{\alpha \alpha'} 
\over eV^2}\right]
\left[ {L \over km}\right] 
\left[ {GeV \over E_{\nu}}\right] \sqrt{f}.
\label{delta}
\end{equation}
The quantity $f$ contains the matter effects. 
For oscillations in vacuum, 
$f = 1$. In general, $f$ is given by,
\begin{equation}
f = 1 + a^2.
\label{f}
\end{equation}
The quantity $a$ is related to the effective 
potential describing the coherent forward scattering
of the neutrino with the background medium.
In the case of ordinary - sterile neutrino
oscillations, the effective potential
includes the effects of both charged and neutral current
interactions.  For $\alpha = e$, i.e. $\nu_e - \nu'_e$ 
oscillations\cite{ref},
\begin{equation}
a = - {\sqrt{2}G_F (2N_e - N_n) E_{\nu}\over \delta m^2_{e e'}},
\label{a}
\end{equation} 
where $G_F$ is the Fermi constant and $N_e$ and $N_n$ are the 
number densities of electrons and neutrons of the 
medium through which the neutrinos propagate. Note that
equal number densities for electrons and protons has 
been assumed in Eq.(\ref{a}).
For neutrino propagation through the Earth, $N_e = N_p \simeq N_n$
where $N_p$ is the number density of protons. 
Hence $N_e$ is related to the density $\rho$  
and the proton mass $m_p$ by $N_e \simeq 0.5\rho/m_p$.

Of course, in a detailed analysis the probability
$P(\nu_{\alpha} \to \nu_{\alpha}, L, E_{\nu})$ must be averaged over
$L$ and $E_{\nu}$ with an appropriate weighting factor,
which takes into account the energy spectrum of the neutrinos,
the cross section and so on.
Such an analysis is quite difficult without detailed knowledge
of the experiments (for example, the lepton detection efficiency 
function is not given in Ref.\cite{kam}) and is most easily
performed by the experimentalists themselves.
However, much can still be learned without doing a rigorous
analysis.

We define the useful quantity $D_{\alpha \alpha'}$ by the distance 
for which $|\delta_m| = \pi/4$, which corresponds to a survival
probability of $1/2$ (if $f \simeq 1$). From Eq.(\ref{delta}),
\begin{equation}
D_{\alpha \alpha'} \simeq  \left[ {E_{\nu} \over GeV} \right]
\left[ {eV^2 \over |\delta m^2_{\alpha \alpha'}|}\right]
\left[{1 \over \sqrt{f}}\right]{\pi \over 4}{1 \over 1.27} 
\ {\rm km.}
\label{D}
\end{equation}
For distances $L \stackrel{<}{\sim} D_{\alpha \alpha'}/2$, 
the oscillation length is too large to have a significant effect.
For $L \stackrel{>}{\sim} D_{\alpha \alpha'}$, the oscillations are 
significant, and should deplete the number of neutrinos by 
a factor of about 2 (if $f \simeq 1$) after suitable averaging 
is performed.  In the intermediate regime, $D_{\alpha \alpha'}/2
\stackrel{<}{\sim} L \stackrel{<}{\sim} D_{\alpha
\alpha'}$, an $L-$ dependent depletion occurs.

In the absense of $\nu_e$ oscillations, the zenith angle
dependent multi GeV atmospheric
neutrino data suggest that the muon neutrino oscillates
maximally with $|\delta m^2_{\mu \mu'}| \sim 10^{-2} \ eV^2$.
In terms of the parameter $D_{\mu \mu'}$, 
this corresponds to
$D_{\mu \mu'} \sim 40 \  km$ for $E_{\nu} \sim 0.6\ GeV$ 
(sub GeV data) and $D_{\mu \mu'} \simeq 400\ km$ 
for $E_{\nu} \sim 6 \ GeV$ (multi GeV data).  
For the sub GeV data, therefore, there should not be
much zenith angle dependence since most of the neutrinos travel 
distances greater than $40\ km$.
There may be some effect for neutrinos coming close to vertically
down, however
at low energies the correlation between the lepton direction
and the incident neutrino direction is quite weak.

Qualitatively it is clear that the atmospheric neutrino
anomaly suggests that if the electron
neutrino oscillates maximally with a sterile neutrino then
$D_{e e'} \gg D_{\mu \mu'}$,  otherwise there would not be 
a significant decrease in the ratio $R$.
The effect of $\nu_e - \nu'_e$ oscillations will be
most important for the low energy neutrino data, since
for low energies, the length $D_{ee'}$ is reduced.
Let us define the quantities $R_{\pm}$ where
$R_+$ is the contribution to $R$ from neutrinos with
$\cos\theta > 0$ and $R_-$ is the contribution to $R$ from
neutrinos with $\cos \theta < 0$ ($\theta$ is the zenith angle,
with $\theta = 0$ corresponding to downward travelling neutrinos).
Neutrinos with $\cos \theta > 0$ travel through the
atmosphere (where matter effects can be neglected)
for distances $ 20 \stackrel{<}{\sim} L/km \stackrel{<}{\sim} 500$,
while neutrinos with $\cos \theta < 0$ travel in matter for
distances $500 \stackrel{<}{\sim} L/km \stackrel{<}{\sim} 13000$.
In the absense of oscillations, $R_+$ and $R_-$ should
each contain approximately half of the interactions.

Note that matter effects will be important when
$a \stackrel{>}{\sim} 1$. 
Furthermore, for $a \stackrel{>}{\sim} \sqrt{2}$, the effects 
of the $\nu_e - \nu'_e$ oscillations become suppressed and
can be approximately neglected. From Eq.(\ref{a}), 
the quantity $a$ can be expressed as
\begin{equation}
a \simeq 1.5 \left[{10^{-4} \ eV^2 \over \delta m^2_{ee'}}\right]
\left[ {\rho \over 4\ g/cm^3}\right]
\left[{E_{\nu} \over GeV}\right].
\label{chau}
\end{equation}
Assuming that $\rho \simeq 4 \ g/cm^3$\cite{qq}, the condition 
$a \stackrel{>}{\sim} \sqrt{2}$ implies that
\begin{equation}
|\delta m^2_{ee'}| \stackrel{<}{\sim} 7 \times 10^{-5}\ eV^2 \ 
{\rm for}\ E \simeq 0.6 \ GeV.
\label{eee}
\end{equation}
Thus for the above range of parameters the matter effects
ensure that the $\nu_e - \nu'_e$ oscillations can be approximately
neglected for $\nu_e - \nu_e'$ oscillations through the Earth.
The $\nu_e - \nu'_e$ oscillations during propagation through the 
atmosphere can also be neglected if
$|\delta m^2_{ee'}| \stackrel{<}{\sim} 7 \times 10^{-5} \ eV^2$, 
since for this range of 
$|\delta m^2_{ee'}|$, $D_{ee'} 
\stackrel{>}{\sim} 4000 \ km \gg 500 \ km$ . 

Observe that for $|\delta m^2_{ee'}|
\stackrel{>}{\sim} 7 \times 10^{-5} \ eV^2$, 
the length $D_{ee'} \stackrel{<}{\sim} 
4000 \ km$ for the sub GeV neutrinos. Thus, for $\delta m^2_{ee'}$
in this range, the $\nu_e - \nu'_e$ oscillations
will be important and will reduce the number of electron neutrinos. 
This will increase $\langle R\rangle$, 
where the brackets $\langle ... \rangle$ 
denote the average over all zenith angles. 
As $|\delta m^2_{ee'}|$ is increased, 
$D_{ee'}$ decreases and $\langle R\rangle$ increases (and matter
effects quickly become negligible for the sub-GeV neutrinos).
For $E_{\nu} \simeq 0.6 \ GeV$, the value
$|\delta m^2_{ee'}| \simeq 10^{-3}\
eV^2$ corresponds to $D_{ee'} \simeq 500 \ km$ which
is the distance to the horizon 
(that is, the $\theta = 90$ deg line).
In order to obtain insight into the increase of 
$\langle R\rangle$ with 
$|\delta m^2_{ee'}|$, it is useful to explicitly calculate 
$\langle R\rangle$ 
for this point. This is because neutrinos coming from the hemisphere
with $0 \le \theta \le \pi/2$ travel distances less than
$D_{ee'}$, whereas those from the $\pi/2 \le \theta \le \pi$
hemisphere travel distances greater than $D_{ee'}$.
So, $R_-$ should be approximately equal to the standard model
value since both $\nu_e$ and $\nu_{\mu}$ fluxes are
reduced by a factor of 2, while $R_+$ will be about half of
the standard model value. This leads to $\langle R\rangle \simeq 0.67$
for $|\delta m^2_{ee'}| \simeq 10^{-3}\ eV^2$.
Clearly $|\delta m^2_{ee'}| \stackrel{>}{\sim}
10^{-3} \ eV^2$ implies that $\langle R \rangle \stackrel{>}{\sim} 
0.67$, with $R_- \simeq 1 $ and $0.5 \stackrel{<}{\sim} R_+ 
\stackrel{<}{\sim} 1$. For some value of $|\delta m^2_{ee'}|$,
$\langle R\rangle$ becomes so large that
it is disfavoured by the data that actually
suggest an anomaly. Note that the current laboratory bound 
$|\delta m^2_{ee'}| < 7.5 \times 10^{-3}\ eV^2$\cite{exp} is 
probably in the range where $\langle R\rangle$ crosses over
into disfavoured values.
In summary then, for $\delta m^2_{ee'}$ in the range
\begin{equation}
|\delta m^2_{ee'}| \stackrel{>}{\sim} 7 \times 10^{-5}  
\ eV^2
\label{hmm}
\end{equation} 
the effects of the maximal $\nu_e - \nu'_e$ oscillations should be
significant for the low energy data. 
The effect of the $\nu_e - \nu'_e$ 
oscillations with $\delta m^2_{ee'}$ in the range 
Eq.(\ref{hmm}) will be to increase $\langle R \rangle$.  

As well as increasing $\langle R \rangle$, the $\nu_e - \nu'_e$ 
oscillations will also make the flux of 
electron neutrinos zenith angle 
dependent for the low energy data.
The zenith angle dependence should manifest itself by an 
{\it increase} in
$R$ for decreasing values of $\cos \theta$. 
Such a result would not be expected
if the anomaly is interpreted assuming 
only $\nu_{\mu} - \nu'_{\mu}$
oscillations and this should provide a distinctive signature 
for $\nu_e - \nu_e'$ oscillations. However,
since the angular correletion between the neutrinos 
and the produced
charged leptons is quite poor at low 
energies (r.m.s. $\sim 60^o$), this
effect will be quite difficult to measure.  
Although no evidence for 
zenith angle dependence in the low energy data
has been found by the existing experiments, the 
sensitivity should be greatly improved in the near future 
with the data expected from Superkamiokande.
In this context, we remark that a sensitive way to test for
$\nu_e - \nu'_e$ oscillations is to compare the measured number
of electron events with $\cos \Theta > 0$, $n_e^+$, with the
number of electron events with $\cos \Theta < 0$, $n^-_e$.
(Here $\Theta$ is the zenith angle of the detected electron).
If electron neutrino oscillations are negligible then
$R_e \equiv n^+_e/n_e^- \simeq 1$. If $\nu_e$ 
oscillations occur with
$\delta m^2_{ee'}$ in the range Eq.(\ref{hmm}), then
$R_e > 1$. Note that $R_e$ should be almost free of systematic
uncertainties. With the improved statistics expected from 
the superKamiokande experiment it should be possible to measure
$R_e$ quite accurately. Thus the 
hypothesis that $\nu_e$ oscillates
maximally with a sterile $\nu_e'$ can be tested, provided 
that $\delta m^2_{ee'}$ is in the range Eq.(\ref{hmm}).

Finally, it only remains to comment on the 
impact of $\nu_e - \nu'_e$ oscillations on the multi-GeV data.
For $|\delta m^2_{ee'}| \stackrel{<}{\sim} 
7 \times 10^{-4} \ eV^2$, $a \stackrel{>}{\sim} \sqrt{2}$ 
for $E_{\nu} \sim 6 \ GeV$ [see Eq.(\ref{chau})], which 
means that the matter effects will suppress 
the $\nu_e - \nu'_e$ oscillations through the Earth.
Thus, for much of the parameter space of interest, Eq.(\ref{hmm}),
$\nu_e - \nu'_e$ oscillations will not modify the expectations
for the multi-GeV atmospheric neutrino data.
There may be some effects for $\nu_e - \nu'_e$ oscillations 
with $\delta m^2_{ee'}$ in the range
$|\delta m^2_{ee'}|
\stackrel{>}{\sim} 7 \times 10^{-4} \ eV^2$.
The effect of $\delta m^2_{ee'}$ in this
range would be to
increase $R$ for $\cos \theta \sim -1$. 

In conclusion we have examined the implications of the
hypothesis that both the $\nu_{\mu}$ and $\nu_e$ neutrinos 
are maximally mixed with sterile partners
for the atmospheric neutrino experiments.
The assumption of maximal mixing means that the oscillations
can be described by just two parameters, $\delta m^2_{ee'}$ and
$\delta m^2_{\mu \mu'}$.
We have shown that the maximal $\nu_e - \nu'_e$ oscillations should
not significantly affect the atmospheric neutrino experiments if
$\delta m^2_{ee'}$ is in the range
$|\delta m^2_{ee'}| \stackrel{<}{\sim} 7 \times 10^{-5}\ eV^2$.
For $\delta m^2_{ee'}$ in the remaining range,
$|\delta m^2_{ee'}| \stackrel{>}{\sim} 7 \times 10^{-5} \ eV^2$,
the effects of the $\nu_e - \nu'_e$ oscillations will be significant.
They should lead to the distinctive signature of an
increasing value of $R$ for lower values of $\cos \theta$ for
the low-energy neutrino sample.
This prediction may be tested by the new data expected
from Superkamiokande.  A related effect 
is that the value of $R$ averaged over zenith
angle, $\langle R \rangle$,
should be somewhat higher than the expected value 
in the absense of $\nu_e - \nu'_e$
oscillations (again for the low-energy sample).  
Further data, especially from Superkamiokande, should clarify 
these issues.

\vskip 1cm
\noindent
{\large \bf Acknowledgements}
\vskip 0.5cm
\noindent
This work was supported by the Australian Research Council.
R.F. is an Australian Research Fellow.

\newpage

\vskip 1cm
\noindent
{\large \bf Table Captions}
\vskip 0.5cm
\noindent
Table 1: Summary of the current low energy atmospheric neutrino
experiments ($E_{\nu} \sim 0.6$ GeV). Note however that the 
Frejus result includes both contained and semicontained events. 

\newpage
\vskip 2cm
\noindent
{\large \bf Table 1}
\vskip 1.4cm
\tabskip=0pt \halign to \hsize{
\vrule#\tabskip=0pt plus 1fil\strut&
\hfil#\hfil& \vrule#& \hfil#\hfil&
\tabskip=0pt\vrule#\cr
\noalign{\hrule}
&Experiment&& $\langle R \rangle$ &\cr
\noalign{\hrule}
&Kamiokande\cite{kam}&&$0.60^{+0.06}_{-0.05}(stat.) \pm 
0.05 (syst.)$&\cr
&IMB\cite{imb}&&$0.54\pm 0.05 (stat.) \pm 0.12 (syst.) $&\cr
&Soudan2\cite{sou}&&$0.72\pm 0.19(stat.)^{+0.05}_{-0.07}(syst.)$&\cr
&Frejus\cite{frejus}&&$1.00\pm 0.15(stat.) \pm 0.08(syst.)$&\cr
&Nusex\cite{nusex}&&$1.04 ^{+0.28}_{-0.32}$&\cr
\noalign{\hrule}
}


\begin{thebibliography}{99}
\bibitem{qq}
The density of the Earth in the mantle (which is about $3000 \
km$ in depth) varies from about $3$ to $5.5 \ g/cm^3$.
The density of the core is significantly larger, in the range
$10$ to $13 \ g/cm^3$. The increase in density of the core can
suppress neutrino oscillations during the neutrino propagation
through the core even when $a \stackrel{<}{\sim} 1$ for neutrinos
propagating through the mantle. However in Eq.(\ref{eee}),
it is justified to take $\rho$ as the density of the mantle 
(that is $\rho \simeq 4 \ g/cm^3$)
since the neutrinos which pass through the core must also pass 
through at least $6000\ km$ of mantle. In the region of interest
where $a \stackrel{<}{\sim} \sqrt{2}$ in the mantle, $D_{ee'}
\stackrel{<}{\sim} 4000\ km \le 6000\ km$.
Thus any suppression of the neutrino oscillations while
the neutrinos are propagating through the core does not matter.
\end{thebibliography}
\end{document}